# Classification of COVID-19 anomalous diffusion driven by mean squared displacement

Yingjie Liang, Peiyao Guan, Shuhong Wang, Lin Qiu

College of Mechanics and Materials, Hohai University, Nanjing, Jiangsu 211100, China

**Corresponding authors**: Yingjie Liang, Email: liangyj@hhu.edu.cn

Lin Qiu, Email: qiulinstyle@126.com

**Abstract:** In this study, we classify the COVID-19 anomalous diffusion in two categories of countries based on the mean squared displacement (MSD) of daily new cases, which includes the top four countries and four randomly selected countries in terms of the total cases. The COVID-19 diffusion is a stochastic process, and the daily new cases are regarded as the displacements of diffusive particles. The diffusion environment of COVID-19 in each country is heterogeneous, in which the underlying dynamic process is anomalous diffusion. The calculated MSD is a power law function of time, and the power law exponent is not a constant but varies with time. The power law exponents are estimated by using the bi-exponential model and the long short-term memory network (LSTM). The bi-exponential model frequently use in magnetic resonance imaging (MRI) can quantify the power law exponent and make an easy prediction. The LSTM network has much better accuracy than the bi-exponential model in predicting the power law exponent. The LSTM network is more flexible and preferred to predict the power law exponent, which is independent on the unique mathematical formula. The diffusion process of COVID-19 can be classified based on the power law exponent. More specific evaluation and suggestion can be proposed and submitted to the government in order to control the COVID-19 diffusion.

# 1. Introduction

On 29 July 2020, more than 16 million cases of coronavirus disease 2019 (COVID-19) had been confirmed including 0.66 million deaths and 10 million recovered [1]. The COVID-19 pandemic is global threat to our societies, and needs to be tackled by the effort from different communities across all fields and disciplines. Scientists have fought against the COVID-19 pandemic from different perspectives [2-4]. One of the hot research topics is mathematical and physical modeling to explain and predict the occurrence and the temporal evolution mechanism of the COVID-19 spread [5-9]. In this study, we try to quantify the COVID-19 dynamics from diffusion. For each country, the common phenomena are that total cases increase and the daily new cases fluctuate with increasing time based on different conditions, such as spread environment, control measurements, and incubation period. To interpret these two behaviors from a statistical mechanics perspective, we consider the COVID-19 dynamic is particles diffusion within the country in any regions. The COVID-19 diffusion is considered a stochastic process. The diffusion environment of COVID-19 in each country is heterogeneous, and the underlying dynamic process is not Brownian motion but anomalous diffusion.

Mean squared displacement (MSD) provides a common denominator for anomalous diffusion, which is no longer a linear function of time in anomalous diffusion as [10]:

$$<x^2(t)> \sim t^\alpha, \tag{1}$$

when $\alpha=1$, the governing stochastic process is the classical Brownian motion, while if $\alpha<1$, the process is sub-diffusion; and if $\alpha>1$, the process is super-diffusion. Anomalous diffusion has been well studied by numerous popular anomalous diffusion models [11-12], specifically, the continuous time random walk (CTRW) models [13], the fractional derivative models [14], the Hausdorff derivative models [15], and the nonlinear partial differential equation (PDE) models, i.e., time or space dependent diffusion models [16]. By determining the value of the power law exponent in Eq. (1), the COVID-19 diffusion can be classified.

It is known that for a diffusion process with one single scale, the power law exponent in Eq. (1) is a constant, which can fully capture the statistical properties without transient phenomenon [17]. But for the diffusion process with temporal or spatial multi-scales, the power law exponent in Eq. (1) will not be a constant, but fluctuates with increasing time or space [18-19]. The physical meaning of the variable values of $\alpha$ in Eq. (1) has been well investigated in the context of the variable order or distributed order fractional derivative diffusion models [20-21]. Here we will focus on the pattern of the power law exponent for specific COVID-19 anomalous diffusion processes in different countries. Based on the existing data and results, it is known that the COVID-19 diffusion process in any country is very complicated, and the corresponding daily new cases change dramatically in different periods. Additionally, the diffusion process is infected by many different measurements, which should contain transients with increasing time. Thus, the power law exponent in Eq. (1) for each country should not be a constant but vary in different time scales $t^{\alpha(t)}$.

In this study, the COVID-19 diffusion processes in two categories of countries are investigated. The first category includes US, Brazil, Russia and India, which are the top four countries with total cases of COVID-19. The second category, selected randomly, includes Spain, UK, Canada and Singapore. For each country, the daily new cases are considered as the displacements of diffusive particles. The MSDs and the corresponding power law exponents in Eq. (1) can be calculated, which fluctuate with increasing days. To predict the power law exponent, two strategies are employed including the bi-exponential model [22], which is frequently used in magnetic resonance imaging (MRI) and can quantify the anomalous diffusion effected by fast and slow diffusion processes [23-24], and the LSTM network [25-26], a popular deep learning way, is used to fit and predict the power law exponents. The diffusion processes of COVID-19 in the selected countries will be deeply investigated and classified based on the MSD.

## 2. Methods

*2.1 Data*

In this study, the COVID-19 diffusion processes in two categories of countries are investigated from the period February 15th to July 3rd. The data of daily new cases are from the website https://www.worldometers.info/coronavirus/#countries. The first category includes US, Brazil, Russia and India, which are the top four countries in total cases of COVID-19. The second category includes Spain, UK, Canada and Singapore, which are randomly selected. For each country, the daily new cases are considered as the displacements of diffusive particles. The MSD and the corresponding power law exponents in Eq. (1) can be calculated, which fluctuate with increasing days. The unit of time used to calculate the power law exponent is *s*. The bi-exponential model and the LSTM network are used to fit and predict the power law exponents, more details of the methods can be found in Appendix.

*2.2 Fitting and Analysis*

The parameters in the bi-exponential model for each country are estimated using the curving fitting toolbox in MATLAB. Based on the deep learning toolbox in MATLAB, the LSTM network is trained in terms of the 80% data length of the power law exponents for each country. To predict the power law exponent, the number of hidden units is 288. In the training options, the solver is 'adam' and the epochs for training are 250 [25]. To prevent the gradients from exploding, the gradient threshold is set to 1. The root mean square errors (RMSEs) of the simulation results are also calculated for the bi-exponential model and the LSTM network.

### 3. Applications and Discussion

*3.1 COVID-19 diffusion in US, Brazil, Russia and India*

The daily new cases of US, Brazil, Russia and India are given in Fig. 1(a) from the period February 15th to July 3rd. The corresponding MSDs are calculated and shown in Fig. 1(b). Fig. 1(c) provides the fitted values of the power law exponent $\alpha$ in MSD by using the bi-exponential model. Fig. 1(d) illustrates the predicted values of the power law exponent $\alpha$ by using the LSTM network. The parameters in the bi-exponential model are given in Table A1 in Appendix. Table 1 provides the RMSEs for the bi-exponential model and the LSTM network.

Table 1. RMSEs of the bi-exponential model and the long short-term memory network for US, Brazil, Russia and India

| RMSE | US | Brazil | Russia | India |
| --- | --- | --- | --- | --- |
| Bi-exponential model | 0.0998 | 0.0204 | 0.0245 | 0.0386 |
| LSTM | 0.0016 | 0.0037 | 0.0033 | 0.0062 |

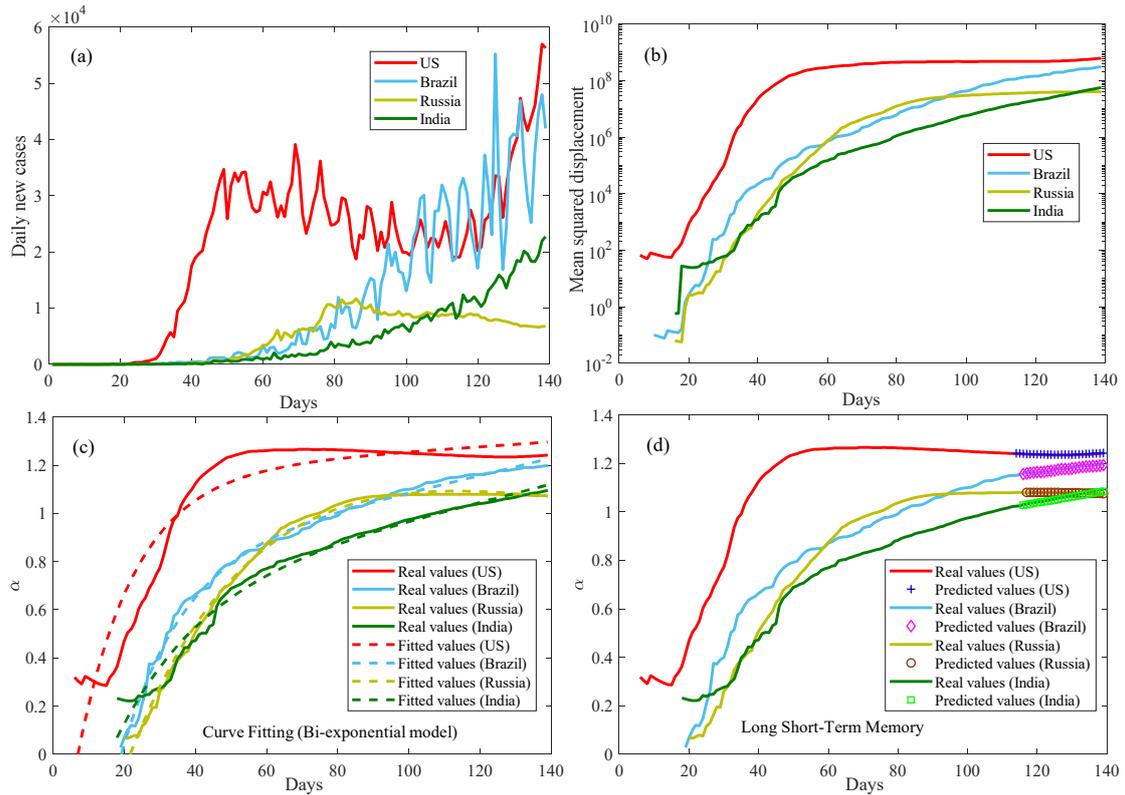

Fig. 1 Plots of (a) daily new cases, (b) mean squared displacement, (c) curve fitting of the power law exponent by the bi-exponential model, and (d) the predicted power law exponent by the long short-term memory network for US, Brazil, Russia and India.

Except for Russia, the current states of the daily new cases for US, Brazil and India have clear increasing trend, and the current values of $\alpha$ for the three countries are larger than 1 conforming that the diffusion processes of COVID-19 are super-diffusion. Based on the results in Fig. 1(c), for US, Brazil and India the diffusion processes transit from sub-diffusion to super-diffusion, but for Russia, the COVID-19 diffusion process transits from sub-diffusion to normal diffusion, which clarifies the fluctuations of daily new cases given in Fig. 1(a). The increase patterns for the MSD in Fig. 1(b) and the values of $\alpha$ in Fig. 1(c) are similar with the day increases.

More specifically, it can be observed from Fig. 1(a) that the daily new cases of US exceed ten thousand and increase dramatically from about the 40th day. The underlying values of $\alpha$ are larger than 1 indicating that the COVID-19 diffusion in US is super-diffusion, and it reaches to a flat curve and lasts a very long period. For Brazil and India, the increasing trends are clear for the daily new cases, and the values of $\alpha$ increase with a similar pattern. For Russia, the daily new cases reach to a high point in the 80th day, then after that it decreases slowly with the day increase. The corresponding MSD and the values of $\alpha$ have a stable state, and maintain a plat increase.

In Fig. 1(c), except for US, the fitted curves for the values of $\alpha$ are almost consistent with the patterns in the real cases. For US, the pattern for the increasing values of $\alpha$ cannot be very accurately estimated by using the bi-exponential model, because of the two periods of peak and valley in the curve of daily new cases. Based on the results in Fig. 1(c), the bi-exponential model can quantify the values of $\alpha$ and makes an easy prediction using an explicit mathematical formula. But it is easy to estimate the patterns of $\alpha$, and its accuracy is not very good, e.g., the case for US, to estimate the values of $\alpha$ with increasing days based on the RMSE in Table 1.

It can be seen from Fig. 1(d) that the predicted results for the 20% data length of the power law exponents given by the LSTM network can well match the real values of the power law exponent $\alpha$ in terms of the 80% data length of the power law exponents. The RMSE for each country in Table 1 is much smaller than that of the bi-exponential model. Compared with the bi-exponential model, the LSTM network is more flexible and preferred to predict the values of $\alpha$, which do not depend on the unique mathematical formula. Thus, the LSTM network is also an alternative strategy to predict the patterns of the power law exponent $\alpha$ for the four countries.

*3.2 COVID-19 diffusion in Spain, UK, Canada and Singapore*

The daily new cases of Spain, UK, Canada and Singapore are given in Fig. 2(a) from the period February 15th to July 3rd. The MSDs of the daily new cases are shown in Fig. 2(b). Fig. 2(c) gives the fitted values of the power law exponent $\alpha$ by using the bi-exponential model, and its parameters are provided in Table A2 in Appendix. The predicted values of the power law exponent $\alpha$ are displayed in Fig. 2(d) by using the LSTM network. Table 2 provides the RMSEs for the bi-exponential model and the LSTM network.

In Fig. 2(a), the daily new cases of Spain, UK, Canada and Singapore are less than ten thousand for the highest peak of the curves and less than two thousand in the current state, which is different from the results of the four countries in Fig. 2(a). The curves of the corresponding MSDs for the daily new cases in Fig. 2(b) reach to a stable state quickly and last for a very long period. The values of $\alpha$ in Fig. 2(c) indicate that the COVID-19 diffusion processes for Canada and Singapore are sub-diffusion. For UK it transits from sub-diffusion to normal diffusion, but for Spain it transits from sub-diffusion to super-diffusion, then transits to normal diffusion, which quantifies the trends of fluctuations in the daily new cases given in Fig. 2(a). In Spain and UK, more large increases exist in the daily new cases compared with those of Canada and Singapore, and the corresponding COVID-19 diffusion processes belong to a slight sup-diffusion when the values of $\alpha$ are larger but almost equal to 1. For the MSDs in Fig. 2(b) and the values of $\alpha$ in Fig. 2(c), they have similar patterns with the day increases, and the curves of $\alpha$ for the four countries have a slight decay trend at the present state, which are consistent with the low increase in the fluctuations of the daily new cases.

In Fig. 2(c), the fitted curves for the values of $\alpha$ by using the bi-exponential model have similar shapes and can generally capture the patterns in the real values expect for large fluctuations in the case of Singapore. Based on the results in Figs. 2(c) and 3(c), the bi-exponential model with four parameters is capable to describe the COVID-19 diffusion with different patterns and can be used to estimate the values of $\alpha$ in a simple mathematical form. In Fig. 2(d), it is also found that the LSTM network can predict the values of $\alpha$ for the 20% data length very well based on the 80% data length of the real values of $\alpha$. And the RMSE for each country in Table 2 is very small, which is better than the bi-exponential model from the perspective of accuracy. Based on the results in Figs. 2(d) and 3(d), the LSTM can be accepted to predict the power law exponent. Based on the values of $\alpha$, the diffusion process of COVID-19 can be classified and more specific evaluation and suggestion can be proposed and submitted to the government in order to control COVID-19 diffusion.

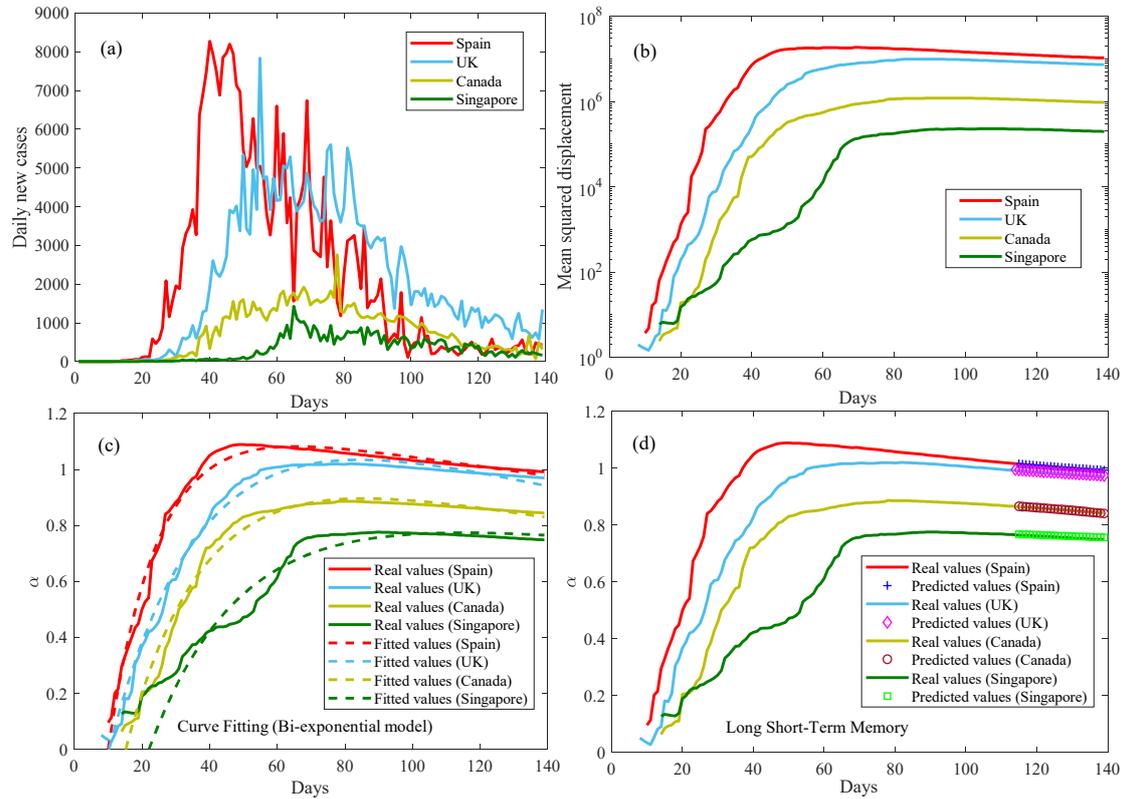

Fig. 2 Plots of (a) daily new cases, (b) mean squared displacement, (c) curve fitting of the power law exponent by the bi-exponential model, and (d) the predicted power law exponent by the long short-term memory network for Spain, UK, Canada and Singapore.

Table 2.  RMSEs of the bi-exponential model and the long short-term memory network for Spain, UK, Canada and Singapore

| RMSE | Spain | UK | Canada | Singapore |
|---|---|---|---|---|
| Bi-exponential model | 0.0277 | 0.0302 | 0.0311 | 0.0376 |
| LSTM | 0.0010 | 0.0041 | 0.0016 | 0.0058 |

**4. Concluding remarks**

In this study, we classify the COVID-19 anomalous diffusion in two categories of countries based on the MSD of daily new cases. The first category includes US, Brazil, Russia and India, which are the top four countries in total cases of COVID-19. The second category includes Spain, UK, Canada and Singapore, which are randomly selected. The COVID-19 diffusion is considered as a stochastic process, and the corresponding daily new cases are equal to the displacements of diffusive particles. The diffusion environment of COVID-19 in one country is heterogeneous, and the underlying it is anomalous diffusion. Based on the data of daily new cases, the corresponding MSD can be calculated. The MSD is a power law function of time, and the power law exponent is not constant, which varies with the increasing time. The values of $\alpha$ classify the diffusion types for different specific time periods.

To predict the patterns of the power law exponent $\alpha$, the bi-exponential model in a traditional simple mathematical form frequently used in MRI and the long short-term memory network as a popular recurrent neural network method are used.

The results show that the bi-exponential model can quantify the values of $\alpha$ and makes an easy prediction using an explicit mathematical formula. The accuracy of the bi-exponential model is not very good, and the physical mechanism from the fast and slow diffusion processes should be further considered in the next step. The LSTM network can predict the values of $\alpha$ for the 20% data length very well based on the 80% data length of the real values, which has better accuracy than the bi-exponential model. And the LSTM network is more flexible and preferred to predict the values of $\alpha$, which do not depend on the unique mathematical formula.

Based on the values of $\alpha$, the diffusion process of COVID-19 can be classified and more specific evaluation and suggestion will be proposed and submitted to the government in order to control the COVID-19 diffusion. The effect of the measurements, e.g., lock down, can be quantified based on the values of $\alpha$, which should approach zero for the sub-diffusion of COVID-19 controlled in a very stable state and lasts a very long time.

**Acknowledgements**

The work described in this paper was supported by the National Natural Science Foundation of China (No. 11702085), and the Fundamental Research Funds for the Central Universities (No. B210202098).

**References**
[1] https://www.worldometers.info/coronavirus/#countries
[2] Song P, Wang L, Zhou Y, He J, Zhu B, Wang F, Tang L, Eisenberg M. An epidemiological forecast model and software assessing interventions on

COVID-19 epidemic in China. medRxiv 2020; 2020: 20029421.

[3] Boldog P, Tekeli T, Vizi Z, Denes A, Bartha F, Rost G. Risk assessment of novel coronavirus COVID-19 outbreaks outside China. J Clin Med 2020; 9: 571.

[4] Li X, Zhao X, Sun Y. The lockdown of Hubei province causing different transmission dynamics of the novel coronavirus (2019-nCov) in Wuhan and Beijing. medRxiv 2020; 2020: 20021477.

[5] Zhang Y, Yu X, Sun H, Tick G, Wei W, Jin B. Applicability of time fractional derivative models for simulating the dynamics and mitigation scenarios of COVID-19. Chaos Soliton Fract 2020; 138: 109959.

[6] Lu Z, Yu Y, Chen Y, Ren G, Xu C, Wang S, Yin Z. A fractional-order SEIHDR model for COVID-19 with inter-city networked coupling effects. arXiv: 2020; 2004.12308.

[7] Zhou T, Liu Q, Yang Z, Liao J, Yang K, Bai W, Lu X, Zhang W. Preliminary prediction of the basic reproduction number of the Wuhan novel coronavirus 2019-nCoV. J Evidence-Based Med 2020; 13: 3-7.

[8] Benvenuto D, Giovanetti M, Vassallo L, Angeletti S, Ciccozzi M. Application of the ARIMA model on the COVID-2019 epidemic dataset. Data Brief 2020; 29: 105340.

[9] Rodriguez O, Conde-Gutierrez R, Hernandez-Javier A. Modeling and prediction of COVID-19 in Mexico applying mathematical and computational models. Chaos Soliton Fract 2020; 138: 109946.

[10] Eliazar I, Shlesinger M. Fractional motions. Phys Rep 2013; 527: 101-29.

[11] Metzler R, Klafter J. The random walk's guide to anomalous diffusion: a fractional dynamics approach. Phys Rep 2000; 339: 1-77.

[12] Sokolov I. Models of anomalous diffusion in crowded environments. Soft Matter 2012; 8: 9043-52.

[13] Cortis A, Knudby C. A continuous time random walk approach to transient flow in heterogeneous porous media. Water Resour Res 2006; 42: W10201.

[14] Liang Y, Chen W, Akpa B, Neuberger T, Webb A, Magin R. Using spectral and cumulative spectral entropy to classify anomalous diffusion in Sephadex™ gels. Comput Math Appl 2017; 73: 765-74.

[15] Liang Y, Ye A, Chen W, Gatto R, Colon-Perez L, Mareci T, Magin R. A fractal derivative model for the characterization of anomalous diffusion in magnetic resonance imaging. Commun Nonlinear Sci 2016; 39: 529-37.

[16] Su N, Sander G, Liu F. Similarity solutions for solute transport in fractal porous media using a time-and-scale-dependent dispersivity. Appl Math Model 2005; 29: 852-70.

[17] Ingo C, Magin R, Colon-Perez L, Triplett W, Mareci T. On random walks and entropy in diffusion-weighted magnetic resonance imaging studies of neural tissue. Magnet Reson Med 2014; 71: 617-27.

[18] Weeks E, Weitz D. Subdiffusion and the cage effect studied near the colloidal glass transition. Chem Phys 2001; 284: 361-7.

[19] Liang Y, Chen W, Xu W, Sun H. Distributed order Hausdorff derivative diffusion model to characterize non-Fickian diffusion in porous media. Commun Nonlinear

Sci 2019; 70: 384-93.

[20] Sun H, Zhang Y, Chen W, Reeves D. Use of a variable-index fractional-derivative model to capture transient dispersion in heterogeneous media. J Contam Hydrol 2014; 157: 47-58.

[21] Su N, Nelson P, Connor S. The distributed-order fractional diffusion-wave equation of groundwater flow: Theory and application to pumping and slug tests. J Hydrol 2015; 529: 1262-73.

[22] Marcon M, Keller D, Wurnig M, Weiger M, Kenkel D, Eberhardt C, Eberli D, Boss A. Separation of collagen-bound and porous bone-water longitudinal relaxation in mice using a segmented inversion recovery zero-echo-time sequence. Magn Reson Med 2017; 77: 1909-15.

[23] https://www.mathworks.com/discovery/lstm.html

[24] Szilvia A, Mihaly A, et al. Bi-exponential diffusion signal decay in normal appearing white matter of multiple sclerosis. Magn Reson Imaging 2013; 31: 286-95.

[25] Sun G, Jiang C, Wang X, Yang X. Short-term building load forecast based on a data-mining feature selection and LSTM-RNN method. IEEJ T Electr Electr 2020; 15: 1002-10.

[26] Luo X, Li D, Yang Y, Zhang S. Spatiotemporal traffic flow prediction with KNN and LSTM. J Adv Transport 2019; 2019: 4145353.

**Appendix**

**A1. The bi-exponential model**

The bi-exponential model generalizes the classical exponential model with four parameters to be determined, which is expressed as

$$\alpha(t) = a\exp(bt) + c\exp(dt), \qquad (A.1)$$

where $a$, $b$, $c$ and $d$ are constants. The power law exponent $\alpha$ is not a constant, and fluctuates with time. Eq. (2) is used to estimate the values of power law exponent $\alpha$ with a simple mathematical form. Eq. (2) is provided in the curving fitting toolbox in MATLAB [23]. It is noted that the bi-exponential model is often used to describe signal decay in the field of MRI, in which the weighted parameters satisfy $a + c = 1$, and $b$ and $d$ are the slow and fast diffusion coefficient respectively [24]. Mono-exponential model is not suitable to the COVID-19 anomalous diffusion, but is feasible for normal diffusion. Table A1 and A2 provided the parameters in the bi-exponential model for the COVID-19 anomalous diffusion in the eight countries.

Table A1. Parameters in the bi-exponential model for US, Brazil, Russia and India

| Countries | $a$ | $b$ | $c$ | $d$ |
|---|---|---|---|---|
| US | 1.17 | 8.89e-9 | -1.82 | -7.25e-7 |
| Brazil | 0.81 | 3.43e-8 | -2.55 | -6.54e-7 |
| Russia | 1.85 | -3.91e-8 | -2.98 | -2.93e-7 |
| India | 0.72 | 3.68e-8 | -1.41 | -4.56e-7 |

Table A2. Parameters in the bi-exponential model for Spain, UK, Canada and Singapore

| Countries | $a$ | $b$ | $c$ | $d$ |
|---|---|---|---|---|
| Spain | 1.239 | -1.956e-8 | -2.424 | -7.917e-7 |
| UK | 1.466 | -3.56e-8 | -2.066 | -4.201e-7 |
| Canada | 1.162 | -2.751e-8 | -2.178 | -5.045e-7 |
| Singapore | 0.933 | -1.528e-8 | -2.072 | -4.329e-7 |

**A2. The long short-term memory network**

The long short-term memory (LSTM) network is a kind of recurrent neural network (RNN), which has ability to deal with the long-term temporal correlations in time series. The topology LSTM with one cell is shown in Fig. A1, which is adapted from the Ref. [25].

The LSTM is conducted for the time series $X(t) = [x_1, x_2, ..., x_L]$, $L$ is the data length, by using the following steps [24]:

Input gates

$$I_i^t = \sum_{l=1}^{L} \omega_{li} x_l^t + \sum_{h=1}^{H} \omega_{hi} v_h^{t-1} + \sum_{m=1}^{M} \omega_{mi} s_c^{t-1}, \tag{A. 2}$$

$$v_i^t = \sigma(I_i^t), \tag{A. 3}$$

Forget gates

$$I_f^t = \sum_{l=1}^{L} \omega_{lf} x_l^t + \sum_{h=1}^{H} \omega_{hf} v_h^{t-1} + \sum_{m=1}^{M} \omega_{mf} s_c^{t-1}, \tag{A. 4}$$

$$v_f^t = \sigma(I_f^t), \tag{A. 5}$$

Cells

$$I_e^t = \sum_{l=1}^{L} \omega_{le} x_l^t + \sum_{h=1}^{H} \omega_{he} v_h^{t-1}, \tag{A. 6}$$

$$s_c^t = v_f^t s_c^{t-1} + v_i^t g(I_e^t), \tag{A. 7}$$

Output gates

$$I_o^t = \sum_{l=1}^{L} \omega_{lo} x_l^t + \sum_{h=1}^{H} \omega_{ho} v_h^{t-1} + \sum_{m=1}^{M} \omega_{mo} s_c^{t-1}, \tag{A. 8}$$

$$v_o^t = \sigma(I_o^t), \tag{A. 9}$$

Cell outputs

$$v_c^t = v_o^t h(s_c^t), \tag{A. 10}$$

where $H$ is the number of hidden layer, $M$ is the number of memory cells, $\omega_{ij}$ is the weight of the connection from unit $i$ to unit $j$, $I_j^t$ is the network input to unit $j$ at time $t$, $v_j^t$ is the value after activation function in the same unit, $s_c^t$ is the state of cell at time $t$, $\sigma$ is the activation function of the gates, $g$ and $h$ are respectively the activation functions of the input and output of cell. More details can be found in [25-26].

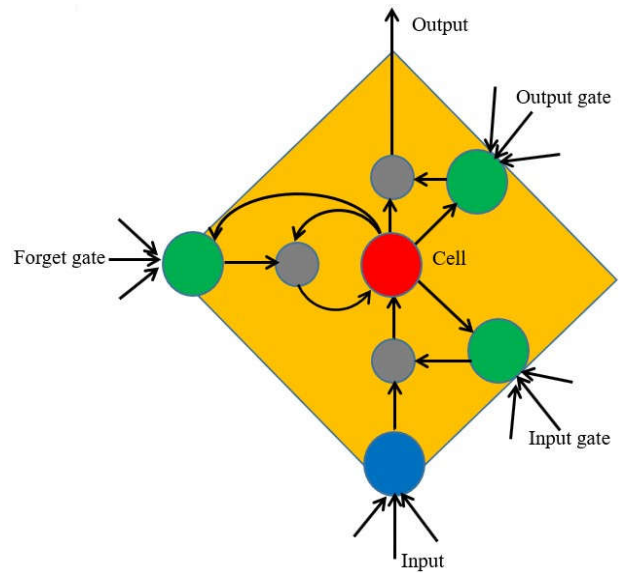

Fig. A1 Topology of LSTM network with one cell [25].